\documentclass[12pt,preprint]{aastex}

\shorttitle{CS in M~82 and NGC~4038} \shortauthors{Bayet et al.}

\begin{document}


\title{Tracing high density gas in M~82 and NGC~4038.}


\author{E. Bayet\altaffilmark{1}, C. Lintott\altaffilmark{3},
  S. Viti\altaffilmark{1},
  J. Mart\'in-Pintado\altaffilmark{2}, ,
  S. Mart\'in\altaffilmark{4}, D.A. Williams\altaffilmark{1}
  and J.M.C. Rawlings\altaffilmark{1}}

\email{eb@star.ucl.ac.uk}


\altaffiltext{1}{Department of Physics and Astronomy, University
  College London, Gower Street, London WC1E 6BT, UK.}
\altaffiltext{2}{Departamento de Astrofisica Molecular e Infrarroja -
  Instituto de Estructura de la Materia-CSIC, C Serrano 121,
  E-28006 Madrid, Spain}
\altaffiltext{3}{Oxford Astrophysics, The Denys Wilkinson Building,
Keble Road, Oxford OX1 3RH, United Kingdom}
\altaffiltext{4}{Harvard
Smithsonian Center for Astrophysics,
  60 Garden Street, Cambridge, MA 02138, USA}


\begin{abstract}

We present the first detection of CS in the Antennae galaxies towards
the NGC~4038 nucleus, as well as the first detections of two high-J
(5-4 and 7-6) CS lines in the center of M~82. The CS(7-6) line in
M~82 shows a profile that is surprisingly different to those of other low-J CS
transitions we observed. This implies the presence of a separate, denser and warmer
molecular gas component. The derived physical properties and the
likely location of the CS(7-6) emission suggests an association with
the supershell in the centre of M~82.
\end{abstract}


\keywords{astrochemistry --- ISM: molecules --- submillimeter ---
galaxies: individual (Antennae, M~82) --- stars: formation}



\section{Introduction}\label{sec:intro}

The molecule CS is a good tracer of dense gas (n(H$_2$)$\geq$ 10$^{5}$-10$^{7}$ cm$^{-3}$) in massive star-forming regions in our own Galaxy (\citealt{Plum92, Plum97,
Andr07}) and in nearby galaxies such as the Magellanic Clouds
\citep{Niko07}, M~51 and Maffei~2 \citep{Pagl95}, NGC~253
\citep{Maue89a, Maue89b, Mart05}, IC~342 and M~82 \citep{Henk85,
Maue89a, Maue89b}. Multi-line studies of CS are crucial for the
determination of the average gas densities in galaxies since CS
transitions have excitation thresholds ranging from 10$^{4}$-10$^{5}$
cm$^{-3}$ for the CS(2-1) line \citep{Bron96} up to $\sim 2\times
10^{7}$ cm$^{-3}$ for the CS(7-6) transition \citep{Plum92}. In this letter, we report in the center of M~82 the first detections of the CS(5-4) and CS(7-6) transitions and the detection of the CS(5-4) line
towards the Antennae galaxies (NGC~4038). We have also re-observed lower-J CS lines in M~82. These two well-known sources were chosen because
interferometric submillimeter/millimeter maps previously obtained
have shown high concentrations of molecular gas. High
resolution $^{12}$CO(1-0) maps indicate that the nearby (D = 13.8Mpc,
see \citealt{Savi04}) Antennae interacting galaxies are likely sites
of rapid high mass star formation \citep{Wils00}. M~82 represents an excellent example of a nearby (D=3.25 Mpc, see
\citealt{Dumk01}) starburst galaxy. Low-J molecular line studies (e.g. \citealt{Fuen06, Seaq06,
Mart06}) have determined its average gas physical parameters. However, the molecular emission from the M~82
nucleus appears to come from multiple gas components. In particular the detection of abundant CH$_{3}$OH \citep{Mart06} and HCN \citep{Brou93} indicate the presence of high density gas in the nucleus. Our detection of the CS(7-6) line not only clearly
confirms the presence of very high density gas but its analysis (see Sect.~\ref{sec:discu}) suggests that it is located in the expanding
superbubble likely to be associated with supernoave remnants \citep{Weis99, Will99, Yao06}.

\section{Observations and results}\label{sec:obs}

The observations of the CS(5-4) line in the Antennae galaxies
(NGC~4038 nucleus) and the CS(7-6) transition in M~82 were performed
during the spring 2007 at the James Clerk Maxwell Telescope (JCMT). We used a position switching mode under medium
weather conditions ($\tau_{225}$ = 0.16). For observing the CS(7-6)
transition ($\nu$ = 342.883GHz) towards the center of M~82, we used
the heterodyne HARP-B multi-beam receiver and the ACSIS digital
autocorrelation spectrometers with a bandwidth of 250MHz. The
receiver noise temperature (single sideband mode) for the HARP-B
central pixel (corresponding to the center of M~82) was $\sim$ 344K.
The telescope main beam efficiency and half power beam width (HPBW)
under this configuration were 0.63 and $\sim$ 14$''$, respectively.
The CS(5-4) line ($\nu$= 244.936GHz) has been observed in NGC~4038
and in M~82 using the receivers RxA3 and A3 of the JCMT, respectively,
again with the ACSIS 250MHz bandwidth and DAS backends, respectively.
The receiver noise temperature (double-side band mode) was 240-450K.
The telescope main beam efficiency and HPBW were 0.69 and $\sim$
20$''$, respectively. For both M~82 and NGC~4038, the pointing and
calibration were performed carefully on planets (Mars and Jupiter)
and on evolved stars. The pointing error was estimated to be $\leq$
2$''$ for NGC~4038 whereas is was larger for M~82 ($<$ 8-10$''$).

In the case of M~82, we have also detected the CS(3-2) ($\nu$ = 146.969GHz), the CS(2-1)($\nu$ = 97.980GHz) and the CS(4-3) ($\nu$ = 195.954GHz) 
lines with the IRAM-30m in wobbler switching mode on June 1997 and
 on May 2008. 
As backend we used a 512 x 1 MHz or the 1024 x 4 MHz filterband. For 
the CS(3-2) line, the system noise temperature was 300K and the
pointing error 2-4$''$ whereas the CS(2-1) and CS(4-3) lines 
show T$_{sys}=$ 281 K and 1655 K respectively with moderate pointing errors, due to unstable
weather conditions. The main beam efficiency and
the HPBW of the telescope were 0.69 and $\sim$ 17$''$ at 147 GHz, 
0.75 and $\sim$ 25$''$ at 97 GHz and 0.63 and $\sim$ 13$''$ at 195 GHz.

\begin{figure}
 \includegraphics[height=12cm]{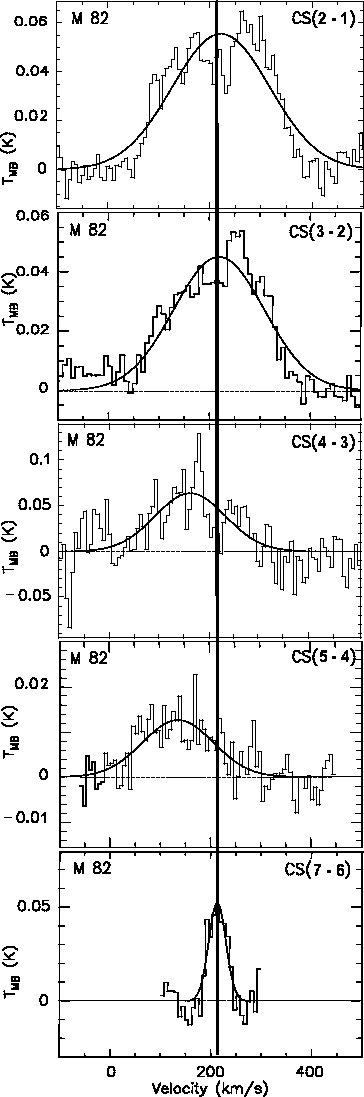}
 \caption{Spectra of the CS(2-1), CS(3-2), CS(4-3), CS(5-4) and CS(7-6) lines
 from top to bottom, respectively, measured towards the center of
 M~82. The observed position corresponds to the center of M~82
 ($\alpha_{\rm J2000}=09^h55^m51.9^s$ and $\delta_{\rm J2000}=+69^\circ40'47''$).
 The CS(2-1) and the CS(4-3) spectra have been smoothed to a common velocity 
 resolution of 6.1kms$^{-1}$ while the CS(3-2), CS(5-4) and CS(7-6) spectra have a velocity 
 resolution of 8.2kms$^{-1}$, $\approx$ 10kms$^{-1}$ and 6.8kms$^{-1}$, respectively. 
 The velocity scale (x axis) is expressed in 
 kms$^{-1}$ units while the temperature scale (y axis) is expressed in main 
 beam temperature units (T$_{MB}$), converted from the antennae temperature 
 via the main beam efficiencies listed in Sect. ~\ref{sec:obs}. The solid
 black lines superimposed onto each spectrum represents the Gaussian
 fit while the vertical black line marks the systemic M~82
 V$_{LSR}$. }\label{fig:1}
\end{figure}

\begin{figure}
 \includegraphics[height=4cm]{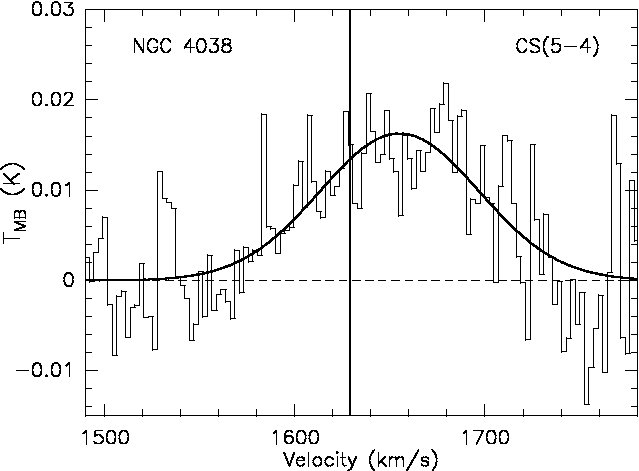}
 \caption{Spectrum of the CS(5-4) line towards NGC~4038. The observed
 position corresponds to the NGC~4038 nucleus ($\alpha_{\rm J2000}=12^h01^m52.8^s$
 and  $\delta_{\rm J2000}=-18^\circ52'05''$). The spectrum has been
 smoothed to a velocity resolution of 2.3 kms$^{-1}$. The solid black
 line superimposed on the spectrum represents the Gaussian fit
 while the vertical black line marks the systemic NGC~4038 V$_{LSR}$.}\label{fig:2}
\end{figure}

All the data have been reduced using either the GILDAS or the Specx
packages, removing the baseline of individual spectra before
averaging them. We fitted Gaussian profiles to the resulting
spectra (Figs. ~\ref{fig:1} and
~\ref{fig:2}). In Table ~\ref{tab:obs}. As shown in Fig. ~\ref{fig:1} and Table 1,
the most striking result is the narrow line width (only 40
kms$^{-1}$) of the CS(7-6) line in M~82, as compared with those 
($\geq$ 200 kms$^{-1}$) of the lower-J CS transitions (see Sect.
~\ref{sec:discu}). Note that both the CS(4-3) and the 
CS(5-4) intensities and linewidths are dependent 
on the baseline we removed and thus suffer from large uncertainties, estimated to be up to 20\% in the total integrated intensity.
However their velocity positions seem robust even if significantly 
shifted from the expected systemic V$_{LSR}$ value of M~82 by a 
factor of 1.3-1.4. Although this characteristic could be explained by
instabilities in the pointing, large pointing differences would produce both a shift in velocity and a narrowing in the linewidth, which are not observed. The uncertainties are therefore attributed to the contamination of the observations by the M~82 South West lobe emission at a velocity of $\sim$100 kms$^{-1}$ (see the CS(2-1) spectrum in Fig. ~\ref{fig:1}).

\section{The physical properties of the high density gas in M~82
and in NGC 4038}\label{sec:ana}

We derive the
following integrated line intensity ratios at the center of M~82 : r$_{32}$ =
$\frac{CS(3-2)}{CS(2-1)}$ = 0.88, r$_{52}$ =
$\frac{CS(5-4)}{CS(2-1)}$ = 0.37. We did not compute the ratio
r$_{72}$ = $\frac{CS(7-6)}{CS(2-1)}$ because the CS(7-6) linewidth is
very different from those found for the other lines and seems to
imply a different gas component than the one responsible for the
lower-J CS transitions (see Sect.~\ref{sec:discu}). Comparing these ratios with those for NGC 253 (see
\citealt{Mart05}), we found substantial differences. The
r$_{32}$ and r$_{52}$ ratios in M~82 are around 3 $\times$ the
values given for the NGC 253-180kms$^{-1}$ component and about 1.8 $\times$
 the values given for the NGC 253-280kms$^{-1}$ component. These
 differences probably arise from different gas excitation
conditions in the M~82 and NGC 253 nuclei (as already suggested
by ~\citealt{Baye04, Baye06}).

\begin{figure}
\includegraphics[height=4cm]{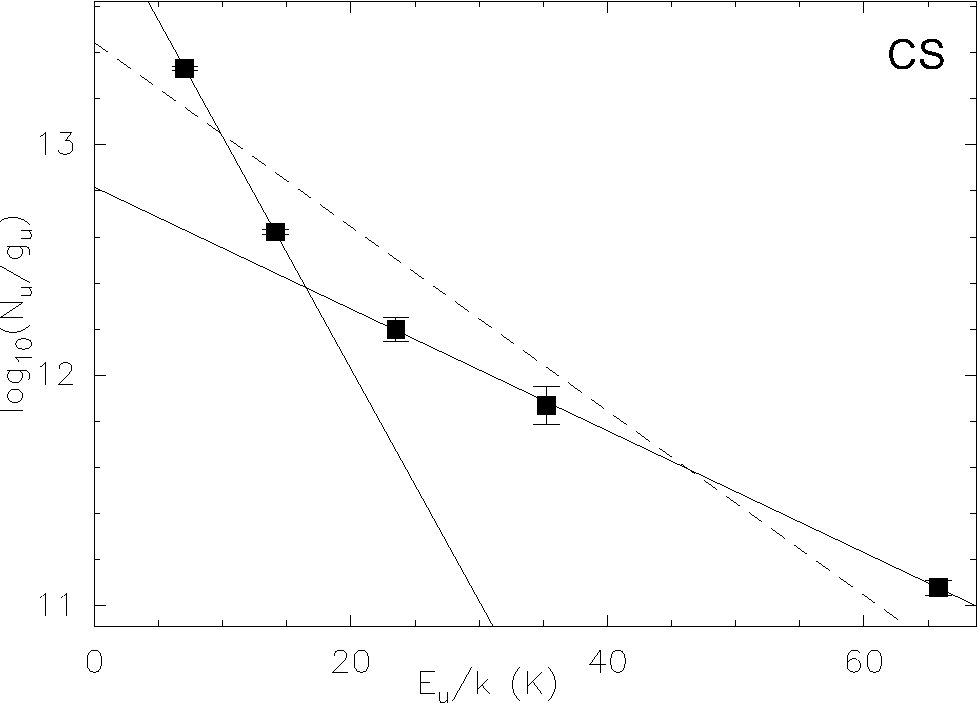}
\caption{Rotation diagram derived from the CS lines towards M~82. The
black lines represent the linear regression fits for two gas components while the dashed line corresponds to one single gas component. The black solid
squares show the data with error bars. These error bars usually represent the main errors in rotational diagrams which actually correspond to those of the integrated intensities (order of 10-20\%). This population diagram has
been corrected for beam dilution effects assuming a source size of 10$''$ for the center of M~82. This value is estimated from the interferometic data presented in \citet{Brou93} and \citet{Garc01}.}\label{fig:3}
\end{figure}

Fig. ~\ref{fig:3} shows the rotation diagram obtained from our observations of CS in M~82, corrected for beam dilution and assuming optically thin emission. As the line profile of the CS(7-6) transition differs significantly from those of the other lines, and as it is expected that the gas in the nucleus of M~82 is not in thermal equilibrium, we favour a fit with two components, as indicated by the methanol observations of \citet{Mart06}. However, due to the large uncertainties of the CS(4-3) and the CS(5-4) lines, it is also possible to make an acceptable fit of the data with one component. Although we have made a 2-component fit, it is likely that there is a continuous change of the excitation temperature. The rotational temperatures for the 2-component fit are 4.3 K and 16.3 K, respectively, while for the single component fit it is 12.4 K. The rotational temperatures derived from the rotational diagram impose a lower limit to the kinetic temperature of the gas. Thus, the detection of the CS(7-6) transition may require a kinetic temperature (T$_{K}$) larger than $\sim$12-15K. To estimate the opacity of the CS lines, observations of isotopes of CS are needed. Unfortunately, no isotope detections have been published so far for this position. The effect of the opacity on the rotation diagram is difficult to predict, but it is likely to result in an increase of the derived rotational temperatures.

To constrain the density and the kinetic temperature traced by the
detection of CS(7-6) line in M~82, we explored LVG models
\citep{Gold74,DeJo75}. Input
parameters ranged from 30 to 100 K for $T_{K}$, from 10$^{5}$cm$^{-3}$ to 10$^{8}$cm$^{-3}$ for n(H$_2$) and from 1$\times 10^{11}$cm$^{-2}$ to 1$\times
10^{15}$cm$^{-2}$ for the CS column density (N(CS)). For each model
of the grid, we fixed the velocity at the CS(7-6) linewidth
value, 40kms$^{-1}$. In these LVG models, the CS collision
partner is He and the collisional rates used are those from
\citet{Liqu07}. We obtained a good agreement between the observed and
modeled CS(7-6) line intensities for a $T_{K}$ range of 65-70 K, a
n(H$_2$) of 1.6$\times 10^{6}$cm$^{-3}$ and a N(CS) of
1.6$\times 10^{14}$cm$^{-2}$. The source-averaged total CS column density (from the observations)
and ranges from 1.3 $\times 10^{13}$cm$^{-2}$- 6.7
$\times 10^{14}$cm$^{-2}$, in agreement with LVG model predictions. Comparing our observations with the radiative transfer model of the
CS molecule developed by \citet{Bene06}, the CS(7-6) observed
integrated line intensity could only be reproduced for T$_{K}$ $>$
50K and n(H$_2$) $\ge$ 10$^{6}$cm$^{-3}$. Indeed, the best fit was
obtained for T$_{K}$=70K and n(H$_2$)=10$^{6}$cm$^{-3}$
(chemical age of 2$\times 10^{3}$yr, see \citealt{Bene06}). This analysis clearly shows that emission from high-J (at least 7-6) CS transitions reveal the presence of very
high density gas in M~82.

For the NGC~4038 nucleus, we were not able to
perform a similar study as no other transitions of CS were observed towards
this source. However, we make a rough estimate the
source-averaged total CS column density from the CS(5-4) line in
NGC~4038 to be N(CS)$=$ 1.8$\times 10^{13}$cm$^{-2}$. In both sources, the CS column densities are consistent
with our model predictions of gas undergoing high-mass star 
formation \citep{Baye08} (see Model 15 
listed in Table 9). For a more detailed analysis, in a forthcoming paper, we shall compare CS predicted line profiles from chemical models with the observational line profiles presented in this Letter.

\section{Discussion}\label{sec:discu}

The angular resolution of the CS(7-6) (beam size of 14") does not
allow us to resolve spatially the emitting region. However we can
estimate its location from the radial velocity and the velocity width
of this line by using the kinematic information provided by
interferometric maps of other molecules. \citet{Brou93} presented a
2" resolution HCN(1-0) map of the south west part of M~82. From the
channel map of the HCN(1-0) emission centered around 213.3
kms$^{-1}$($\pm 40$kms$^{-1}$), it appears that the CS(7-6) emission
may arise from the HCN(1-0) clump located at $\alpha_{\rm
J2000}=09^h55^m52.2^s$ and $\delta_{\rm J2000}=69^\circ40'46.1''$,
within our beam. \citet{Garc01, Garc02} presented high resolution
interferometric HCO and H$^{13}$CO$^{+}$(1-0) maps of M~82. We
re-analyzed the data and found that some of the clumps show narrow
($\leq$ 60 kms$^{-1}$) emission, similar to that of our CS(7-6) line,
confirming the likely location of the CS(7-6) emission at $\approx$
(-5$''$, -2$''$) from the center. This is the location of the
expanding molecular supershell \citep{Weis99, Will99, Yao06}. The high density ($\geq
10^{6}$cm$^{-3}$) and the high temperature ($\approx$ 60-80K) may
arise from the interaction between the expanding supershell and the
ambient gas in M~82.

\begin{table*}
  \caption{Observational parameters.}\label{tab:obs}
  \renewcommand{\footnoterule}{}
  \hspace*{-2cm}
  \begin{tabular}{l c c c c c c c c c }
   \hline
   Source & Line & $\nu$ & Tsys & beam & $\int$(T$_{mb}$ dv)&
   V$_{LSR}$ & $\Delta$V$_{LSR}$ & T$_{peak}$ & rms\\
   & & (GHz) & (K) & size (") &  (Kkms$^{-1}$) &
   (Kkms$^{-1}$) & (Kkms$^{-1}$) & (mK) & (mK)\\
   \hline
   M~82 & CS(2-1) & 97.980 & 281 & 25 & 13.3$\pm$0.3 & 221.9$\pm$2.7 & 225.7$\pm$5.4 & 55.5 & 5.8\\
   & CS(3-2)& 146.969 & 303 & 17 & 11.2$\pm$0.3 & 219.7$\pm$2.8 & 211.3$\pm$6.4 & 50.1 & 4.6\\
   & CS(4-3)& 195.954 & 1655 & 13 & 11.1$\pm$1.3 & 161.0$\pm$9.7 & 165.3$\pm$23.2 & 63.4 & 25.8\\
   & CS(5-4)& 244.936 & 437 & 20 & 4.2$\pm$0.8 & 140.0$\pm$38.0 & 211.2$\pm$44.2 & 15.6 & 10.8\\
   & CS(7-6)& 342.883 & 334 & 14 & 2.2$\pm$0.2 & 213.8$\pm$1.5 & 40.1$\pm$2.8 & 52.5 & 14.1\\
   \hline
   NGC & CS(5-4)& 244.936 & 246 & 20 & 1.7$\pm$0.1 & 1655.0$\pm$4.4 & 98.7$\pm$8.5 & 16.3 & 6.7\\
   4038&&&&&&&&&\\
   \hline
  \end{tabular}
\end{table*}


\vspace*{1cm}

In summary, we have presented the first detection of a high density
tracer (CS) in the Antennae galaxies (NGC~4038 nucleus) as well as
the first detections of two high-J transitions of CS in the center of
M~82. We find that multiple molecular gas components in M~82 are necessary to
explain the observed line intensities. In particular, while low-J CS
lines seem to arise from relatively low density gas ($\sim
10^{5}$cm$^{-3}$), the molecular gas traced by the CS(7-6) line must
be dense ($\sim 10^{7}$cm$^{-3}$) and warm ($\sim$ 70K) and appears to
be associated with the expanding supershell in M~82. The high density and temperature may be due to the interaction
between the expanding supershell and the ambient gas. Similar
multi-line studies for the Antennae galaxies (NGC~4038/39) are
necessary in order to determine the origin of its high density
molecular gas component traced by the CS(5-4) line. In both sources, the CS column densities are consistent with model predictions of gas undergoing high-mass star formation \citep{Baye08}.

Similar extended multi-transition multi-molecule studies
performed on ultra-luminous infra-red galaxies by \citet{Grev06b} and
\citet{Baan08} need to be carried out on nearby sources. In fact,
\citet{Mueh07} presented a detailed study of the para-H$_{2}$CO in
M~82; however this molecule does not trace the dense gas component. Thus, in
order to compare the physical and chemical properties of the very
dense (extragalactic) gas between ULIRGs and nearby sources, CS
observations (from 2-1 to 7-6) of ULIRGs should be performed.

\begin{acknowledgements}
EB acknowledges financial support from the Leverhulme Trust. SV acknowledges individual financial support from STFC AF. This work has been partially supported by the
Spanish Ministerio de Educaci\'on y Ciencia under projects
ESP2004-00665 and ESP2007-65812-C02-01, and ``Comunidad de Madrid"
Government under PRICIT project S-0505/ESP-0237 (ASTROCAM).
\end{acknowledgements}

\bibliographystyle{apj}
\bibliography{references}

\end{document}